# Large perpendicular magnetic anisotropy in epitaxial Fe/MgAl$_2$O$_4$(001) heterostructures

Qingyi Xiang[1,2]*, Ruma Mandal[2], Hiroaki Sukegawa[2], Yukiko K. Takahashi[2] and Seiji Mitani[1,2]*

[1]*Graduate School of Pure and Applied Sciences, University of Tsukuba, Tsukuba, Ibaraki 305-8577, Japan*

[2]*National Institute for Materials Science, Tsukuba, Ibaraki 305-8577, Japan*

*E-mail: XIANG. Qingyi@nims.go.jp, mitani.seiji@nims.go.jp

We investigated perpendicular magnetic anisotropy (PMA) and related properties of epitaxial Fe (0.7 nm)/MgAl$_2$O$_4$(001) heterostructures prepared by electron-beam evaporation. Using an optimized structure, we obtained a large PMA energy ~1 MJ/m$^3$ at room temperature, comparable to that in ultrathin-Fe/MgO(001) heterostructures. Both the PMA energy and saturation magnetization show weak temperature dependence, ensuring wide working temperature in application. The effective magnetic damping constant of the 0.7 nm Fe layer was ~0.02 using time-resolved magneto-optical Kerr effect. This study demonstrates capability of the Fe/MgAl$_2$O$_4$ heterostructure for perpendicular magnetic tunnel junctions, as well as a good agreement with theoretical predictions.

MgAl$_2$O$_4$ is considered a promising alternative barrier material to MgO for magnetic tunnel junctions (MTJs) due to its tunable lattice constant[1,2] and the $\Delta_1$ band preferential transport due to the coherent tunneling effect.[3–6] Especially, a large tunnel magnetoresistance (TMR) ratio[2,7] and improved bias dependence of the TMR ratio[1,8] have been reported in MgAl$_2$O$_4$-based MTJs. In addition to such TMR properties, interface-induced perpendicular magnetic anisotropy (PMA) at an MgAl$_2$O$_4$ interface is a crucial property for applications of perpendicularly magnetized MTJs (p-MTJs). The utilization of perpendicularly magnetized films with a large PMA energy can substantially improve thermal stability of p-MTJs to ensure long data retention for next-generation high-density non-volatile magnetic memories such as spin-transfer-torque magnetoresistive random access memory (STT-MRAM) and magnetoelectric-RAM.[9–19] So far, the largest interface PMA energy density around 1.4 MJ/m$^3$ has been reported in an epitaxial ultrathin-Fe/MgO(001) heterostructure.[20] For MgAl$_2$O$_4$ based epitaxial structures, smaller PMA energy density ~0.4 MJ/m$^3$ has been experimentally reported in Fe/MgAl$_2$O$_4$(001)[21] and Co$_2$FeAl/MgAl$_2$O$_4$(001) heterostructures,[22] where the MgAl$_2$O$_4$ layers were prepared by post-oxidization of an Mg-Al metallic layer. On the other hand, based on a recent theoretical calculation,[23] the areal PMA energy density of ~1.3 mJ/m$^2$ was predicted at an Fe/MgAl$_2$O$_4$(001) interface, which is nearly comparable to that at an Fe/MgO(001) interface (~1.5–1.7 mJ/m$^2$). Interestingly, even the small difference in the PMA densities between Fe/MgAl$_2$O$_4$ and Fe/MgO was clearly interpreted through the second perturbation theory with the orbital resolved densities of states. Therefore, further improvement in the PMA energy of ultrathin-Fe/MgAl$_2$O$_4$(001) interfaces, i.e., observation of the intrinsically large PMA, is expected if a clean interface is obtained by suppressing atomic intermixing and over-oxidation through process optimization. In addition, related magnetic properties of the PMA heterostructures such as magnetic damping and temperature dependence of PMA properties were evaluated: The former determines the switching speed and the current density for MRAM operations, and the latter guarantees the device operation temperature range of p-MTJs.[24,25]

In this study, we investigated magnetic properties of ultrathin-Fe/MgAl$_2$O$_4$ structures fabricated using an electron-beam (EB) evaporation technique to achieve large interface PMA. Through careful tuning of film thicknesses and post-annealing temperatures, an optimized Fe (0.7 nm)/MgAl$_2$O$_4$ interface showed a large PMA energy up to ~1.0 MJ/m$^3$, comparable to the reported value for an Fe (0.7 nm)/MgO (~1.4 MJ/m$^3$).[20] We also found that the PMA energy and saturation magnetization ($M_s$) were not very sensitive to measurement temperature. The effective damping constant was also evaluated to be ~0.02 by time-resolved magneto-optical Kerr effect (TR-MOKE) under high magnetic fields.

Figure 1 (a) shows a schematic design of the multilayer structure to examine the PMA properties at an Fe/MgAl$_2$O$_4$ interface. A fully epitaxial stack of MgO (5 nm)/Cr (30 nm)/Fe ($t_{Fe}$ = 0.7 nm)/MgAl$_2$O$_4$ ($t_{MAO}$ = 2 and 3 nm) was deposited on an MgO(001) substrate by EB evaporation (base pressure ~1×10$^{-8}$ Pa). Before deposition, the substrate was annealed at 800°C to clean its surface, followed by the deposition of the 5 nm MgO seed layer at 450°C. The Cr buffer layer was deposited at 150°C, and then it was post-annealed at 800°C to obtain a flat Cr(001) surface. This post-annealing temperature is critical to obtain a large PMA for an ultrathin Fe layer deposited on the Cr buffer.[20] The temperature conditions for the ultrathin Fe were 150°C and 250°C for growth and post-annealing, respectively, to improve the surface flatness. Then, the MgAl$_2$O$_4$ barrier layer was deposited at 150°C with a ~0.01 nm/s deposition rate from a high-density (98.6% of the theoretical density) sintered MgAl$_2$O$_4$ chip (Ube Material Industries), instead of from an MgAl$_2$O$_4$ substrate as the previous report.[26] The deposited MgAl$_2$O$_4$ barrier was post-annealed at different temperatures (350°C, 400°C, 450°C, and 500°C) to modify the Fe/MgAl$_2$O$_4$ interface conditions. Finally, the 2 nm thick Ru capping layer was sputter-deposited at room temperature (RT). Through the growth process, surface structures and epitaxial growth were *in-situ* monitored by reflection high-energy electron diffraction (RHEED). Magnetic hysteresis loops (*M-H* loops) of the samples were measured using a vibrating sample magnetometer (VSM) at RT and a vibrating sample magnetometer incorporated with superconducting quantum interference device (VSM-SQUID) under temperatures between 100 and 300 K. The ultrafast magnetization dynamics was measured by an all-optical TR-MOKE microscope to evaluate magnetic damping. The 1028 nm fundamental femtosecond laser pulse was used to excite the sample whereas the second-harmonic (wavelength, λ = 515 nm) of the fundamental beam was used to probe the magnetization dynamics by measuring the change in Kerr rotation as a function of time-delay between both pump and probe beams. A variable magnetic field was applied at an angle of 70° with respect to the perpendicular direction of the sample surface.

The RHEED patterns of the sample with an Fe (0.7 nm)/MgAl$_2$O$_4$ (3 nm) were shown in Figs. 1 (b)-(g). As seen in Figs. 1 (d) and (f), the additional sub-streaks indicated by red arrows represent the formation of the c(2×2) reconstructed surface of Cr and Fe, which is believed to improve the surface flatness and consequently the magnitude of PMA of the ultrathin Fe layer when capped with MgO.[20] It is noted that the absence of the c(2×2) structure for Fe was reported in Ref. 18, in contrast to the present study. Besides, as shown in Figs. 1 (b) and (c), the RHEED patterns of the $MgAl_2O_4$ surface after post-annealed at 400°C are similar to those of sputter-deposited $MgAl_2O_4$ on a thick Fe layer.[8] Therefore, the growth of a fully epitaxial structure with (001) orientation was confirmed. The patterns of the $MgAl_2O_4$ surface also indicate that the EB-evaporated $MgAl_2O_4$ in this study has a cation-disordered spinel structure, which ensures the giant TMR effect similar to an MgO barrier.[2,6]

The largest PMA energy density is obtained for Fe (0.7 nm)/$MgAl_2O_4$ (3 nm) with annealing temperature of 400°C. The *M-H* loops of this sample is shown in Fig. 2 (a), where the effective PMA energy density, i.e., $K_{eff}$, was determined by the area enclosed by the in-plane, out-of-plane *M-H* loops, and the *y*-axis (shadow area). The largest $K_{eff}$ reaches ~1.0 MJ/m$^3$, which is comparable to the value (~1.4 MJ/m$^3$) in the previous report for an Fe (0.7 nm)/MgO. Firstly, it should be noted that the $K_{eff}$ observed in this study is more than twice as large as the reported in an Fe (0.7 nm)/$MgAl_2O_4$ (~ 0.4 MJ/m$^3$),[21] where the $MgAl_2O_4$ was prepared by post-plasma-oxidation of an $Mg_{33}Al_{67}$ metallic layer. Secondly, the large PMA which is close to but slightly smaller than that of Fe/MgO is in a good agreement with the theoretical predictions.[23] This fact strongly suggests that the first principles approach describes the mechanism of interface PMA of Fe/oxide correctly. Theoretical calculations also revealed that the over- or under-oxidation at the interface of a ferromagnetic layer and an oxide layer significantly reduces the magnitude of the PMA energy density.[14] Thus, EB-evaporated $MgAl_2O_4$ grown from high-density $MgAl_2O_4$ chips may have improved interface oxidation conditions compared to the post-oxidized $MgAl_2O_4$. It was suggested in Ref. 22 that uniform oxidation of a metal layer is not easy, which tends to cause over-oxidation or under-oxidation at the bottom-side barrier interface depending on the oxidation condition.

By varying the $MgAl_2O_4$ thickness and post-annealing temperature, interface conditions, such as the degree of oxidation, can be tuned.[19] Figure 2 (b) shows $K_{eff}$ as a function of the post-annealing temperature for $t_{MAO}$ = 2 and 3 nm. The samples with $t_{MAO}$ = 3 nm show larger PMA energy density than those with $t_{MAO}$ = 2 nm at all post-annealing temperatures, which may be related to possible variation of oxygen amount near the Fe/$MgAl_2O_4$ interface by increasing the $MgAl_2O_4$ thickness. Moreover, the PMA retains even at 500°C for $t_{MAO}$ = 3 nm, suggesting that the PMA of ultrathin-Fe/$MgAl_2O_4$ is robust enough to endure high-temperature heat treatments during industrial manufacturing.[27]

In addition to the magnitude of $K_{eff}$, weak temperature dependence of $K_{eff}$ is also favorable for practical use of PMA heterostructures. To evaluate the temperature dependence of $K_{eff}$, the $M$-$H$ loops of Fe (0.7 nm)/MgAl$_2$O$_4$ (3 nm) were investigated at different measurement temperatures between 300 K (RT) and 100 K, as shown in Fig. 3 (a). It is found that the shape of the in-plane (hard-axis) loops is significantly temperature dependent. The anisotropy field of the in-plane loops ($H_k$) increases with decreasing temperature, indicating the enhancement of $K_{eff}$ at low temperatures. To analyze the temperature dependence of magnetic properties, we firstly fitted the saturation magnetization $M_s$ by Bloch's law: [28]

$$M_s(T) = M_s(0)\left(1 - \left(\frac{T}{T_c}\right)^{1.5}\right), \quad (1)$$

where $M_s(0)$ is $M_s$ at 0 K, $T$ is the absolute temperature, and $T_c$ is the Curie temperature. The temperature dependence of $M_s$ is plotted in Fig. 3 (b) with the fitting curve using Eq. (1). The fitting results of $T_c$ and $M_s(0)$ are 1227 ± 188 K and is 2.32 ± 0.05 T, respectively. They are close to the values in bulk Fe, i.e., 1043 K and 2.19 T, respectively. Although Bloch's law is not applicable to the temperature range close to $T_c$, the result indicates that $T_c$ of Fe in ultrathin-Fe/MgAl$_2$O$_4$ is not significantly reduced. This is in contrast to the previous reports of ultrathin-Fe on Ag[29] or thin Ni and Co on Cu.[30,31] For $K_{eff}$, we assumed the following simple equation:[32]

$$K_{eff} = \frac{K_i}{t_{Fe}} - 2\pi M_s^2 + K_v, \quad (2)$$

where $K_i$, $-2\pi M_s^2$, and $K_v$ are the interface, shape, and volume anisotropy energy densities, respectively. Here, we assumed $K_v = 0$ for simplicity, and $K_i = t_{Fe}(K_{eff} + 2\pi M_s^2)$ was plotted as a function of $T$ in Fig. 3 (c). The difference in $K_i$ between 100 and 300 K (~2.0 mJ/m$^2$ at 100 K, ~1.7 mJ/m$^2$ at 300 K) appears to be small, compared to that of CoFeB/MgO (~1.9 mJ/m$^2$ at 100 K, ~1.45 mJ/m$^2$ at 300 K),[33] which may be attributed to the high $T_c$ of the Fe layer. Moreover, we fit the $K_i$ by a power law of $M_s(T)$: [33]

$$K_i(T) = K_i(0)\left(\frac{M_s(T)}{M_s(0)}\right)^\gamma, \quad (3)$$

where the $K_i(0)$ is $K_i$ at 0 K. The exponent $\gamma$ = 1.91 ± 0.24 obtained by fitting is close to the values reported in CoFeB/MgO (~2.18 and ~2.16).[33,34] It is worth noting that according to the Callen-Callen law for uniaxial anisotropy, the exponent $\gamma$ = 3 is expected; i.e., $K(T)/K(0)$ = $(M_s(T)/M_s(0))^3$, where $K$ is the anisotropy energy.[35] A reduced exponent was theoretically predicted in the presence of large spin-orbit coupling (SOC) materials that contribute to the PMA,[36–39] and consistent with experiment results in FePt[40]. However, further systematic investigation is required by talking into consideration of $K_v$ and higher order anisotropy for better understanding.

We also evaluated the damping constant ($\alpha_{eff}$) of the ultrathin Fe layer using TR-MOKE method. Figure 4 (a) shows the oscillatory magnetization precessional signals of the Fe (0.7 nm)/MgAl$_2$O$_4$ (3 nm) sample with varying $\mu_0H$. $\alpha_{eff}$ is determined by fitting the TR-MOKE signal with a phenomenological fitting function: [41]

$$G(t) = Ae^{-tt_1} + B\sin(2\pi ft - \varphi)e^{-\frac{t}{\tau}} + C, \qquad (4)$$

where $f$ corresponds to the precessional resonance frequency, $\tau = \frac{1}{2\pi f \alpha_{eff}}$ is the relaxation time, and $\varphi$ is the initial phase of oscillation. $A$ and $B$ denote the amplitudes of oscillations. $C$ and $t_1$ are the offset and the decay rate of demagnetization, respectively. We obtained $\alpha_{eff}$ = 0.0233, 0.0207, and 0.0238 at $\mu_0H$ = 1.77 T, 1.55 T, and 1.27 T, respectively, as shown in Fig. 4 (b), with the lowest $\alpha_{eff}$ obtained ~0.0207. Here, the $\alpha_{eff}$ is not an intrinsic value and only gives the upper limit of true $\alpha$.[42] Note here that layer- and orbital-resolved electronic structure calculations are powerful tools for understanding interface effect of damping, as well as PMA: it is theoretically predicted that the $\alpha$ of thin Fe films is much larger than of the bulk one, with the interfaces of Fe contributed most.[43] Such an enhancement has been observed in ultrathin Fe deposited on Ag, where damping constant for 0.4 nm film is ~9 times larger than for thick films.[44]

In summary, we prepared epitaxial ultrathin-Fe/MgAl$_2$O$_4$ heterostructures by EB-evaporation. A large PMA energy density up to 1.0 MJ/m$^3$ was obtained for the 0.7 nm-Fe/3 nm-MgAl$_2$O$_4$ heterostructure annealed at 400°C, which is in good agreement with the theoretical predictions. The PMA sustained even after post-annealing at 500°C, and the changes in the $M_s$ and PMA energy between 100 and 300 K were relatively small. In addition, the areal PMA energy density $K_i$ is found to be proportional to nearly the square of $M_s$, suggesting that the induced PMA at the Fe/MgAl$_2$O$_4$ interfaces arises from the strong interface SOC. The lowest effective damping constant was estimated to be 0.0207. This study demonstrates robust interface PMA in the ultrathin-Fe/MgAl$_2$O$_4$ useful for p-MTJ applications.

**Acknowledgments**
This study was partly supported by the ImPACT program of the Council for Science, Technology and Innovation (Cabinet Office, Government of Japan) and JSPS KAKENHI Grant Number 16H06332. Q.X. acknowledges National Institute for Materials Science for the provision of a NIMS Junior Research Assistantship.

## Figure Captions

**Fig. 1.** (a) Schematic illustration of an epitaxial heterostructure. (b)-(g) RHEED patterns taken from a sample of Fe (0.7 nm)/MgAl$_2$O$_4$ (3 nm) annealed at 400°C; (b), (d) and (f) The incident electron beams are along [100] azimuth of MgO (001) substrate and (c), (e) and (g) [110] azimuth. Sub-streaks indicated by red arrows correspond to c(2×2) surface structure.

**Fig. 2.** (a) *M-H* loops at RT for sample of Fe (0.7 nm)/MgAl$_2$O$_4$ (3 nm) annealed at 400°C. Shadow area indicates the effective PMA energy density ($K_{\text{eff}}$). Positive $K_{\text{eff}}$ indicates PMA. (b) Annealing temperature dependence of $K_{\text{eff}}$ for Fe (0.7 nm)/MgAl$_2$O$_4$ (2 or 3 nm).

**Fig. 3.** (a) *M-H* loops under different measurement temperatures. (b) Measurement temperature dependence of $M_s$ and $K_i$. The dash lines are fitting results by Eq. (1) and Eq. (3).

**Fig. 4.** (a) Time-dependent signal (scattered data points) of Fe (0.7 nm)/MgAl$_2$O$_4$ (3 nm) under an external bias magnetic field ($\mu_0 H$) with different strengths and their best fit using Eq. (3) (solid black lines). (b) The calculated effective damping constant $\alpha_{eff}$ as a function of $\mu_0 H$.

Fig.1. (Color Online)

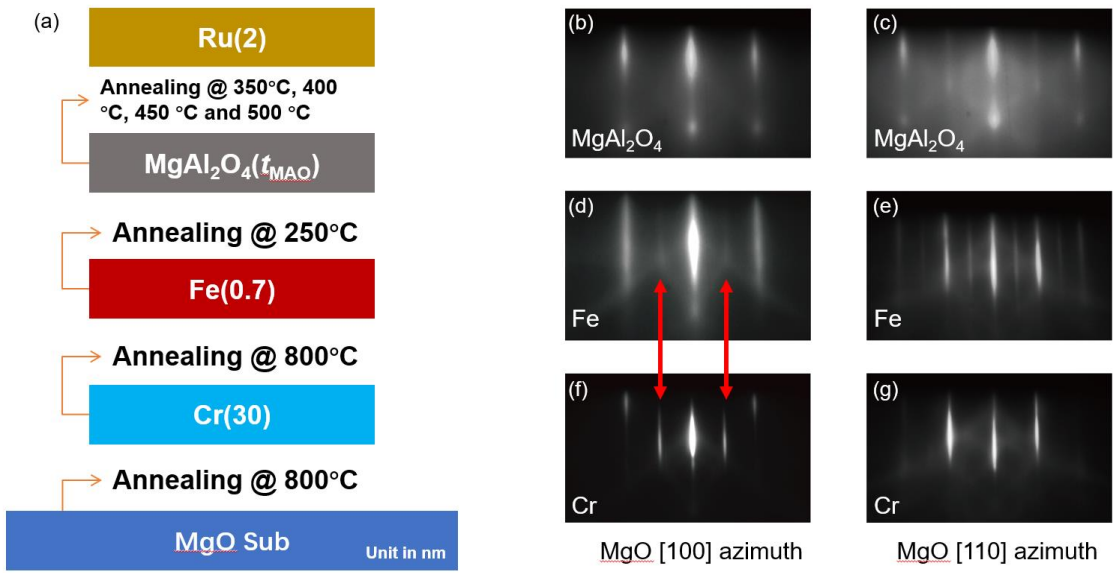

Fig.2. (Color Online)

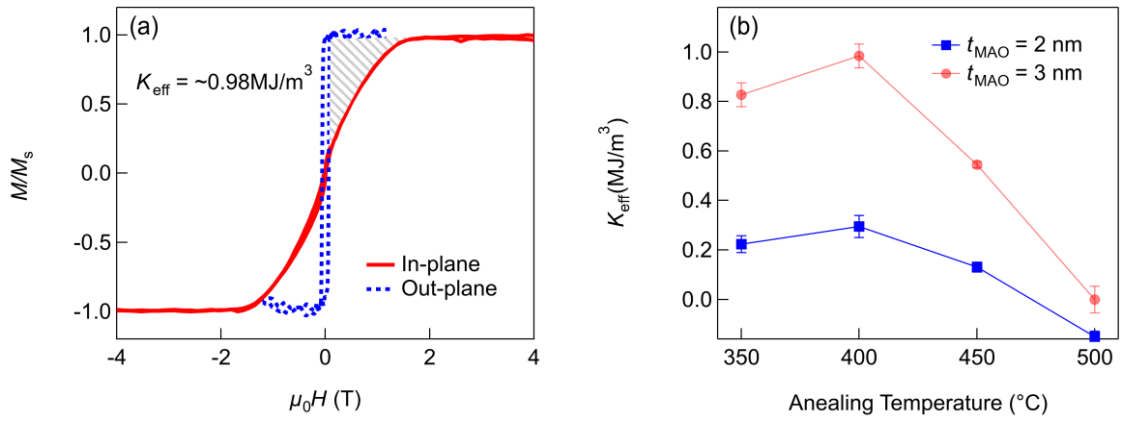

Fig.3. (Color Online)

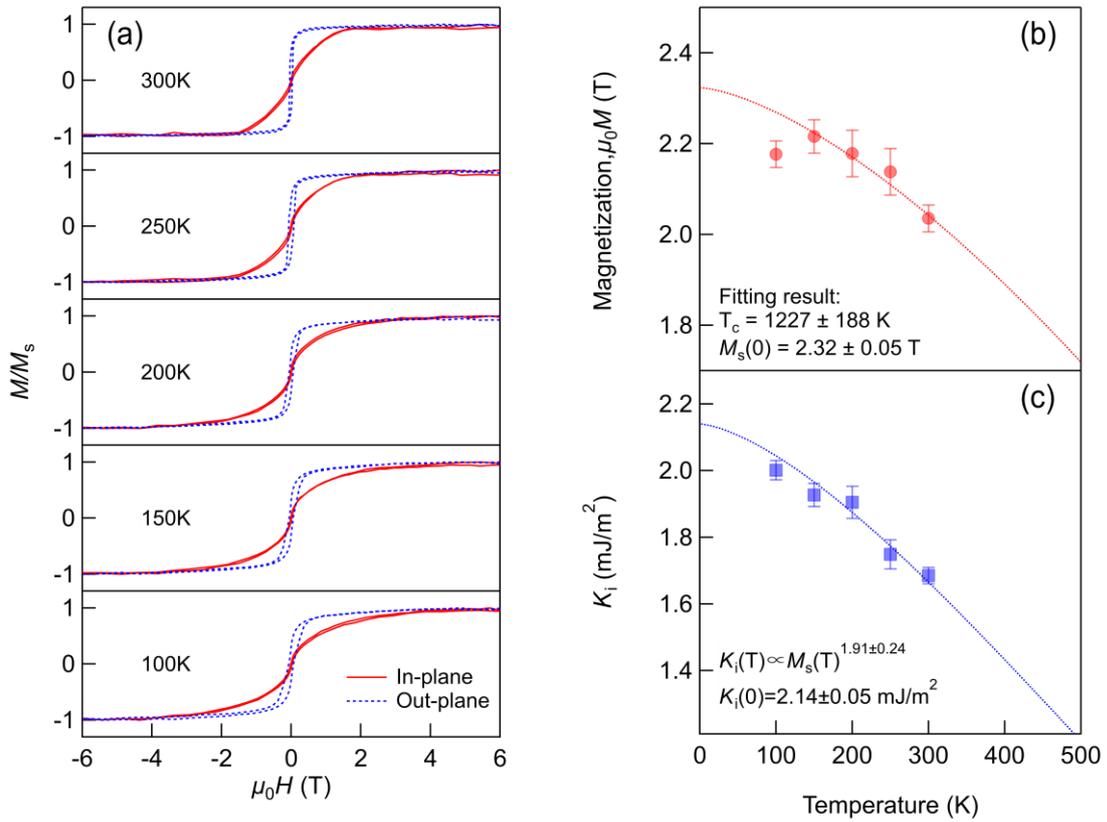

Fig.4. (Color Online)

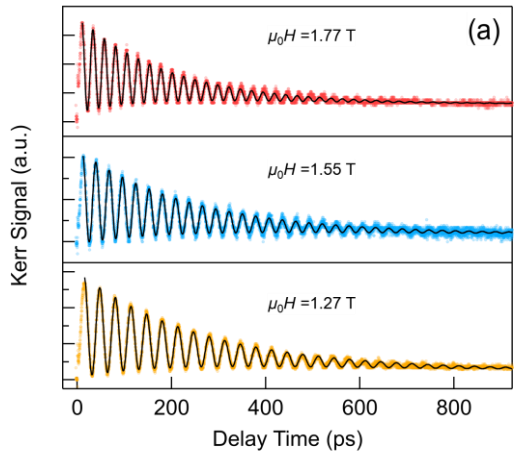 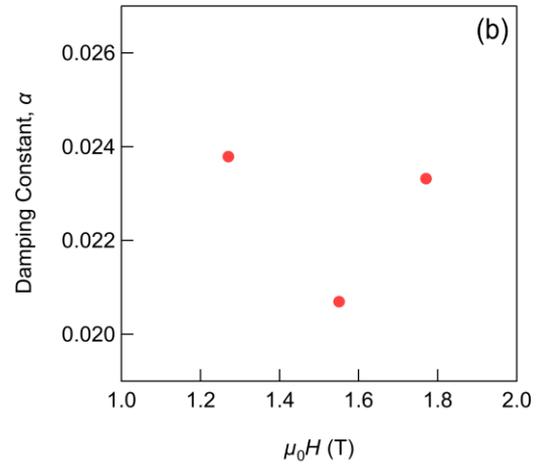